\documentclass[onecolumn,prb,notitlepage,floats,superscriptaddress,amsmath,amssymb]{revtex4-1}

\usepackage[version=3]{mhchem} 
\usepackage{bm}
\usepackage[utf8]{inputenc}
\usepackage[T1]{fontenc}
\usepackage{graphicx}
\usepackage{upgreek}
\usepackage{color}
\usepackage{ulem}
\usepackage{hyperref} 
\hypersetup{colorlinks,citecolor=blue, filecolor=blue ,linkcolor=blue , urlcolor=blue, pdftex}
\usepackage{sidecap}
\usepackage{amssymb}
\usepackage[caption=false]{subfig}

\def \FUW{Institute of Experimental Physics, Faculty of Physics, University of Warsaw, ul. Pasteura 5, 02-093 Warsaw, Poland}
\def \Wroclaw{Department of Semiconductor Materials Engineering, Wrocław University of Science and Technology, ul. Wybrzeże Wyspiańskiego 27, 50-370 Wrocław, Poland}
\def \Watanabe{Research Center for Functional Materials, National Institute for Materials Science, 1-1 Namiki, Tsukuba 305-0044, Japan}
\def \Taniguchi{International Center for Materials Nanoarchitectonics, National Institute for Materials Science, 1-1 Namiki, Tsukuba 305-0044, Japan}

\begin{document}

\title{Temperature induced modulation of resonant Raman scattering in bilayer 2H-MoS$_{2}$ }

\author{Mukul Bhatnagar}
\email{mukul.bhatnagar@fuw.edu.pl}
\affiliation{\FUW}
\author{Tomasz Wo\'zniak}
\affiliation{\Wroclaw}
\author{\L{}ucja Kipczak}
\affiliation{\FUW}
\author{Natalia Zawadzka}
\affiliation{\FUW}
\author{Katarzyna Olkowska-Pucko}
\affiliation{\FUW}
\author{Magdalena Grzeszczyk}
\affiliation{\FUW}
\author{Jan Paw\l{}owski}
\affiliation{\FUW}
\author{Kenji~Watanabe}
\affiliation{\Watanabe}
\author{Takashi Taniguchi}
\affiliation{\Taniguchi}
\author{Adam Babi\'nski}
\affiliation{\FUW}
\author{Maciej R. Molas}
\email{maciej.molas@fuw.edu.pl}
\affiliation{\FUW}

\begin{abstract}
The temperature evolution of the resonant Raman scattering from high-quality bilayer 2H-MoS$_{2}$ encapsulated in hexagonal BN flakes is presented.
The observed resonant Raman scattering spectrum as initiated by the laser energy of 1.96 eV, close to the A excitonic resonance, shows rich and distinct vibrational features that are otherwise not observed in non-resonant scattering. 
The appearance of 1$^{st}$ and 2$^{nd}$ order phonon modes is unambiguously observed in a broad range of temperatures from 5 K to 320 K. 
The spectrum includes the Raman-active modes, $i.e.$ E$_\textrm{1g}^{2}$($\Gamma$) and A$_\textrm{1g}$($\Gamma$) along with their Davydov-split counterparts, $i.e.$ E$_\textrm{1u}$($\Gamma$) and B$_\textrm{1u}$($\Gamma$).
The temperature evolution of the Raman scattering spectrum brings forward key observations, as the integrated intensity profiles of different phonon modes show diverse trends. 
The Raman-active A$_{1g}$($\Gamma$) mode, which dominates the Raman scattering spectrum at $T$=5~K quenches with increasing temperature.
Surprisingly, at room temperature the B$_\textrm{1u}$($\Gamma$) mode, which is infrared-active in the bilayer, is substantially stronger than its nominally Raman-active A$_\textrm{1g}$($\Gamma$) counterpart.

\end{abstract}

\maketitle

\section{Introduction \label{sec:Intro}}
Two-dimensional (2D) semiconducting transition metal dichalcogenides (S-TMDs) have attracted significant attention in the last decade due to the thickness dependent electronic band structure that allows for at-will manipulation of optical and opto-electronic properties~\cite{Lopez-Sanchez2013, Jo2014, Mak2016, Zhu2016, Yan2017}. 
Novel configurations employing monolayers and van der Waals (vdW) heterostructures~\cite{Yu2015, Deng2016, Liao2019} have emerged as promising platforms for the development of cutting-edge technology spanning across a broad spectrum that include but is not limited to quantum information processing~\cite{He2015, Schaibley2016} spintronics,~\cite{Cortes2019, Ciorciaro2020}, nanophotonics~\cite{  Bhatnagar2020, Mennucci2021, Bhatnagar2021} and twistronics~\cite{Carr2017, Chendong}.
From the perspective to study lattice dynamics, non-invasive Raman scattering (RS) spectroscopy  has emerged as a pivotal tool to uncover the physics of vibrational and electronic properties of 2D S-TMDs~\cite{Lee2010, Tonndorf2013, Placidi2015,  Grzeszczyk2016, Zhang2018, Baren2019,  kipczak2020optical}.
In particular, the RS spectrum of the bilayer (BL) MoS$_{2}$, which is one of the simplest prototypes of a multilayer van der Waals structure, uncovers several intriguing features~\cite{Huang2016, Yang2016, Park2016, Lee2017, Lin2018,  Sarkar2019, Debnath2020, Kim2020, Kim2021, Grzeszczyk2021}.
The RS spectrum becomes especially rich under resonant excitation conditions. 
The resonant RS was extensively used to study the characteristics of phonon modes in MoS$_{2}$.
The corresponding RS spectrum comprises in particular double resonance Raman bands, second-order scattering features and Davydov-split pairs~\cite{Sekine1984, GoAa2014, golasa2014resonant, Na2018, Shinde2021}.
The application of electric field~\cite{Lu2017, klein2021electrical} or a variation of temperature~\cite{Gontijo2019, Liv2010,  Kumar2021} become key factors to modulate the electron-phonon coupling within the active 2D layer. 
The latter studies, however, usually do not employ temperatures below those of liquid nitrogen.
To our knowledge, there are no reports on the temperature-dependent resonant RS for BL 2H-MoS$_{2}$ at temperature down to $T$=5~K. 
To fill the gap, we report on the effects observed through temperature-dependent resonant RS in high-quality BL 2H-MoS$_{2}$ encapsulated in hexagonal BN (hBN) flakes. 
\textcolor{red}{It well known in the literature that hBN encapsulation of MoS$_2$ bilayers results in a substantial increase in their quality, manifested in $e.g.$, observation of interlayer excitons, see Refs.~\citenum{Slobodeniuk2019,Paradisanos2020, Grzeszczyk2021}.}
The evolution of the exciton emission energy on temperature as obtained through photoluminescence (PL) measurements allowed us to extract the relative energy difference between the dominant A exciton and the excitation source.
This facilitates studies of the effect of the energy difference on the RS spectrum.
The phonon dispersion, calculated using Density Functional Theory (DFT) reveals the presence of both the Raman- and infrared-active modes that corroborate well with the experiment. 
We demonstrate a temperature-dependent tuning of the integrated intensity profile of the lattice vibrations.
The interplay between the intensity of the Raman-active out-of-plane A$_\textrm{1g}$ mode and its Davydov-split B$_\textrm{1u}$ counterpart is observed.
Surprisingly, the temperature-activated quenching of the A$_\textrm{1g}$ mode with respect to the B$_\textrm{1u}$ mode results in the intensity crossover at 220 K.
The observed results also point to the diverse trends for scattering from different phonon modes from the same points in the Brillouin zone (BZ).

\section{Results and Discussion \label{results}}

\begin{figure*}[h!t]
		\subfloat{}%
		\centering
		\includegraphics[width=1.0\linewidth]{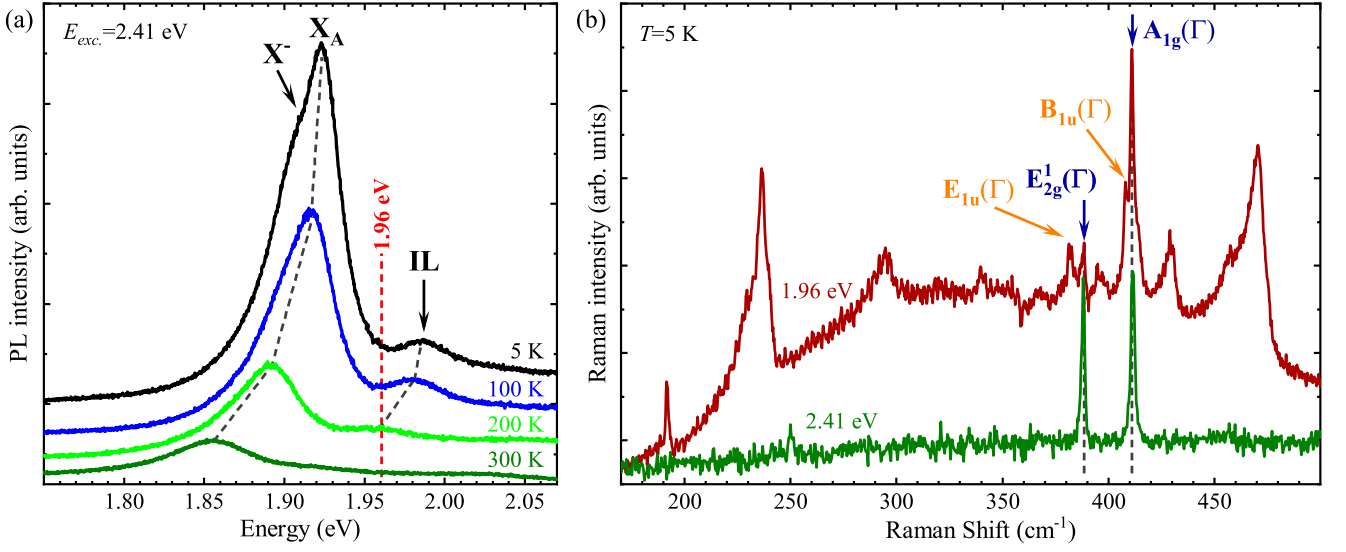}%
    	\caption{\textbf{Photoluminescence and Raman spectra of BL MoS$_{2}$ flake.} (a) Temperature evolution of PL spectra obtained under the 2.41~eV laser excitation. 
    	The black dashed lines denote the energies of the X$^{-}$, X$_{A}$, and IL lines. The vertical red dashed line corresponds to the energy of the 1.96~eV (resonant excitation). 
    	(b) High resolution Raman spectra under non-resonant (2.41 eV) and resonant (1.96 eV) excitation of BL MoS$_{2}$ measured at $T$=5~K. 
    	The main Raman-active modes, $i.e.$ the in-plane E$^1_{2g}$($\Gamma$) and the out-of-plane A$_\textrm{1g}$($\Gamma$), are marked in blue.
    	Their Davydov-split, infrared-active counterparts, $i.e.$ E$_\textrm{1u}$($\Gamma$) and B$_\textrm{1u}$($\Gamma$) are marked in orange. 
    	The vertical dashed lines denote the energetic positions of the modes.}
		\label{fig:fig1}
\end{figure*}
	
It has been well established in the literature that the RS in thin S-TMD layers can be significantly enhanced due to an electron-phonon coupling in the vicinity of excitonic resonances, particularly the so-called A exciton~\cite{Carvalho2015,Molas2017,Zinkiewicz2019,Shinde2021}.
In order to study the effect of resonant conditions on the RS in the MoS$_{2}$ BL, the temperature evolution of the PL spectra is measured in a broad range of temperature from $T$=5~K to 320~K. 
\textcolor{red}{Note that the Raman measurements were carried out on the flat regions of the investigated sample (bubble-free and wrinkles free), see Fig. S1 in the Supplementary Information (SI) for optical and atomic force microscope (AFM) images.
The PL spectra measured at selected temperatures are shown in Fig.~\ref{fig:fig1}(a), while the full set of the measured PL spectra is presented in Fig. S2(a) of the SI.}
The low temperature ($T$=5~K) PL spectrum comprises of three distinct emission lines.
They are ascribed to the negatively charged (X$^{-}$), neutral (X$_{A}$), and interlayer excitons (IL) formed in the vicinity of the A exciton~\cite{Kuemmell2015, Niehues2019, Leisgang2020, Grzeszczyk2021}. 
It can be seen that the temperature increase from 5 K to 300 K leads to the red shift of their emission energies, accompanied by the reduction in the corresponding intensities. 
In particular, the X$_{A}$ line remains dominant of all three resonances up to 300 K, while the X$^{-}$ and IL lines can be resolved only up to about 100 K and 230 K, respectively. 
It is interesting to note that the IL emission energy shifts across the excitation energy (1.96~eV) used for resonant excitation of RS.
To study the effect of temperature on resonant excitation, we analyse the temperature evolution of the relative energy, E$_{L-{X_{A}}}$, defined as the difference between the energy of the resonant excitation (1.96~eV) and of the X$_{A}$ line. 
\textcolor{red}{Fig. S2(a) in the SI} shows the temperature dependence of the X$_{A}$ energy that has been extracted from the PL measurements accompanied by its fitting using the Varshni equation~\cite{Varshni1967}.

\begin{figure*}[h!t]
		\subfloat{}%
		\centering
		\includegraphics[width=1.0\linewidth]{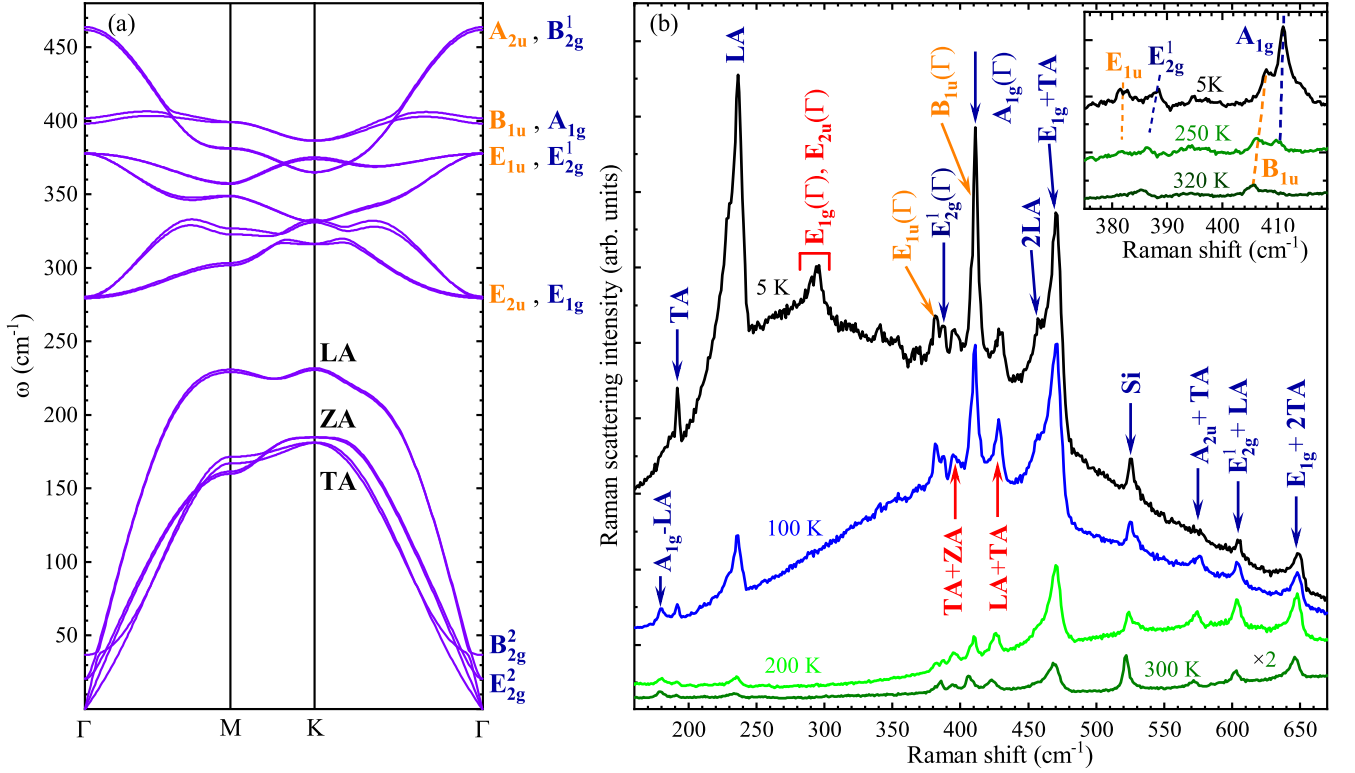}%
		\caption{\textbf{Phonon dispersion and resonant Raman spectrum at different temperature.} 
		(a) The calculated phonon dispersion for BL MoS$_{2}$. 
		The infrared- and Raman-active modes at the $\Gamma$ point of the BZ are marked in orange and blue, respectively. 
		(b) The temperature evolution of the RS spectra measured under resonant condition of excitation (1.96~eV). 
		The phonon modes reported in the literature are marked correspondingly in blue and orange for Raman- and infrared-active vibrations, while the labels of peaks assigned in this work are in red. High resolution Raman spectra  at $T$=5~K, $T$=250~K, and  $T$=320~K revealing Davydov splitting of the in-plane and out-of-plane modes are shown in the inset.}
		\label{fig:fig2}
\end{figure*}

The group theory formalism reveals rich information on crystal symmetry that can be decoded by studying lattice vibrational modes in 2D materials~\cite{scheu2015, Molina2015}. 
The crystal structure for BL belongs to the space group P $\overline{\mbox{3}}$m1, ${\#}$164 (point group D$_{3d}^{3}$)~\cite{nils, Kumar2021} and the normal modes of lattice vibrations at the $\Gamma$ point can be expressed by the following irreducible representation: 3A$_\textrm{1g}$ + 3E$_\textrm{g}$ + 3A$_\textrm{2u}$ + 3E$_\textrm{u}$.
The normal modes in BL correspond to those of the bulk 2H-MoS$_{2}$, although the point groups in both cases are different.
The main rotation axis is sixfold in bulk and not threefold as in BL~\cite{scheu2015}. 
As a result, restricting to optical modes, the normal modes of lattice vibrations at the $\Gamma$ point in bulk can be expressed by the following irreducible representation~\cite{Cai2014}: 
2B$_\textrm{2g}$(A$_\textrm{1g}$)+A$_\textrm{2u}$(A$_\textrm{2u}$)+E$_\textrm{1u}$(E$_\textrm{u}$)+A$_\textrm{1g}$(A$_\textrm{1g}$)+B$_\textrm{1u}$(A$_\textrm{2u}$)+ E$_\textrm{1g}$(E$_\textrm{g}$)+E$_\textrm{2u}$(E$_\textrm{u}$)+2E$_\textrm{2g}$(E$_\textrm{g}$).
The corresponding notation for the BL symmetry is shown in parentheses.
For example, one of the results of the connection between the BL and the bulk is the brightening of bulk-inactive Raman modes in the BL limit previously observed $e.g.$ in thin layers of 2H-MoTe$_{2}$~\cite{GrzeszczykB}. 
Therefore, in the following we will refer to the observed Raman modes using the bulk-related representations, as is usually done in the literature.

The non-resonant and resonant RS spectra measured on BL MoS$_{2}$ at $T$=5~K are shown in Fig.~\ref{fig:fig1}(b).
The increased intensity of phonon scattering at non-standard energies in resonant RS and its complexity compared to non-resonant RS can be related to the proximity of the excitation energy (1.96 eV) to the energy of the A exciton, as was reported earlier~\cite{Carvalho2015,scheu2015}. 
At $T$=5~K, the resonant excitation energy (1.96 eV) is larger by about 34~meV than the X$_{A}$ energy, which corresponds to E$_{L-{X_{A}}}$ of about 276~cm$^{-1}$. 
Consequently, the resonant RS spectrum is superimposed over a background due to the X$_{A}$ emission, which is not present for the non-resonant case. 
The temperature dependence of the observed RS modes will be discussed in the following section.

The calculated phonon dispersion for 2H-MoS$_{2}$ BL is presented in Fig. \ref{fig:fig2}(a)\textcolor{red}{, while the density of phonon states with division into sulfur and molybdenum contributions are shown in Fig. S3 of the SI}. The phonon modes at the $\Gamma$ point of the BZ marked with blue represent Raman active modes, while infrared active lattice vibrations are denoted by orange color. The phonon dispersion is used to investigate in detail the Raman peaks observed in the RS spectra measured as a function of temperature. 
Fig.~\ref{fig:fig2}(b) presents the RS spectra at selected temperatures under resonant excitation of 1.96~eV. 
Let us focus first on the low-temperature ($T$=5~K) RS spectrum. The assignment of most of the Raman peaks, denoted by blue and orange colors, is clear and can be made in reference to the literature~\cite{golasa2014resonant, Chakraborty2013, Lee2015, Shinde2021}. 
The spectrum includes the aforementioned peaks related to both the Raman-active modes, $i.e.$ E$_\textrm{2g}^1$($\Gamma$) and A$_\textrm{1g}$($\Gamma$), as well as their infrared-active counterparts, E$_\textrm{1u}$($\Gamma$) and B$_\textrm{1u}$($\Gamma$). 
\textcolor{red}{The energy difference between the two out-of-plane modes A$_\textrm{1g}$($\Gamma$) and B$_\textrm{1u}$($\Gamma$) is approximately 3 cm$^{-1}$}, corresponding well to the Davydov splitting of the modes reported earlier~\cite{Na2018, Shinde2021}, which supports its attribution. 
The observation of the infrared-active peaks in the RS spectrum is ascribed to the resonant conditions of the laser excitation. 
The A$_\textrm{1g}$(M)-LA(M) mode at 180 cm$^{-1}$ is only observed at a temperature higher than 100~K, which corresponds to its differential character~\cite{golasa2014resonant}. 
As can be appreciated in \ref{fig:fig2}(a), there are no zone-center modes in the energy range below 280 cm$^{-1}$ except for the low-energy shear and breathing modes.
This points out to the zone-edge TA or ZA acoustic phonons most likely around the K point of the BZ as a possible origin of the mode observed at 191 cm$^{-1}$~\cite{Shinde2021}.
Similarly, the mode at 231 cm$^{-1}$ is most likely related to the LA mode at the M and/or K points of the BZ. 
The presence of the zone-edge modes in the RS spectrum is usually related to the disorder in the structure, which leads to the phonon localization and the breaking of Raman momentum conservation in the scattering process\cite{golasaAIP}.
However, their temperature-induced quenching suggests a more complicated process involving the modification of the electron-phonon coupling, which modifies momentum conservation in the resonant RS.
The inspection of the phonon dispersion (see Fig.~\ref{fig:fig2} (a)) also allows us to propose the attribution of the mode at 290 cm$^{-1}$ to the E$_\textrm{2u}$/E$_\textrm{1g}$ zone-centre modes.
The calculations also allow us to propose the assignment of the TA+ZA mode at 397 cm$^{-1}$ and the TA+LA mode at 429 cm$^{-1}$. 
The 2LA branch observed at 462 cm$^{-1}$ is coupled with E$_\textrm{1g}$(M)+TA(M) at 470 cm$^{-1}$~\cite{golasa2014resonant, Lee2015, Kumar2021}.
In the frequency range of 570 cm$^{-1}$ to 650 cm$^{-1}$, the series of Raman-active modes due to two-phonon scattering processes can be distinguished: A$_\textrm{2u}$(M)+TA(M) at 577 cm$^{-1}$, E$_\textrm{2g}^1$(M)+LA(M) at 604~cm$^{-1}$, A$_\textrm{1g}$(M)+LA(M) at 640~cm$^{-1}$, and E$_\textrm{1g}$(M) + 2TA(M) at 650 cm$^{-1}$, which is consistent with previous reports~\cite{golasa2014resonant, Lee2015, Kumar2021}.
The temperature-dependent Raman spectrum under non-resonant conditions is composed only of the two characteristic A$_\textrm{1g}$ and E$_\textrm{2g}^1$ modes throughout the measured energy range \textcolor{red}{(see Fig. S4(a) of the SI), while Fig. S4(b)} presents the variation in the intensity of these two modes with the change in temperature.

\begin{figure*}[!t]
		\subfloat{}%
		\centering
		\includegraphics[width=1.0\linewidth]{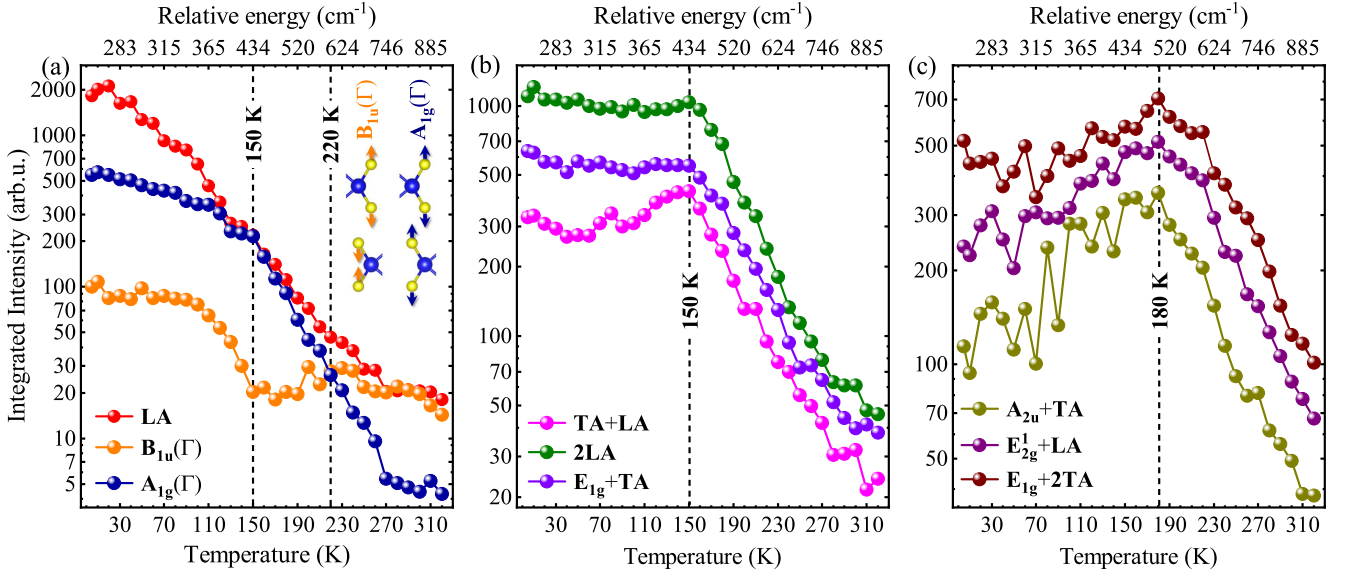}%
		\caption{\textbf{Integrated intensity profile of lattice vibrations extracted from temperature dependent resonant Raman scattering.} 
		(a) The intensity of the sum of the LA-related peak (red) accompanied with the Davydov-split pair of the infrared-active B$_\textrm{1u}$($\Gamma$) (orange) and Raman-active A$_\textrm{1g}$($\Gamma$) (blue) modes as a function of temperature (bottom axis) and of relative E$_{L-{X_{A}}}$ energy (top axis). 
		The inset shows the corresponding B$_\textrm{1u}$($\Gamma$) and A$_\textrm{1g}$($\Gamma$) lattice vibrations that are in colors matched to the intensity curves for the direction of atomic displacements. 
		The corresponding intensity profiles for TA+LA, 2LA, and E$_\textrm{1g}$+TA are shown in panel (b), while for A$_\textrm{2u}^2$+TA, E$^1_\textrm{2g}$+LA and E$_\textrm{1g}$+2TA in panel (c).
		Note the logarithmic scale of the vertical axis.}
		\label{fig:fig3}
\end{figure*}

To gain further insight into the temperature evolution of the observed phonon modes, we performed a detailed analysis of their integrated intensity profiles. 
The specific peaks were fitted with Gaussian functions combined with a linear function to take into account the variation in the background intensity. The obtained intensity profiles of 9 peaks are presented in Fig.~\ref{fig:fig3} as a function of temperature (bottom axis) and relative E$_{L-{X_{A}}}$ energy (top axis). \textcolor{red}{The conversion of the relative energy (eV) into cm$^{-1}$ further facilitates to understand the role of exciton influence on to the vibration energy of the observed Raman modes.}
The profiles are grouped into three panels according to the observed temperature evolution. 
Let us focus on the temperature dependence of the intensity of the LA mode, shown in Fig.~\ref{fig:fig3}(a). 
Its intensity reveals a quick exponential decay of about 2 orders of magnitude in the temperature range from 5 K to 320~K (note logarithmic scale of the vertical axis). 
A similar exponential decay is observed for the A$_\textrm{1g}$($\Gamma$) mode with a decrease of about 2 orders of magnitude from 5 K to 320~K. 
The analogous temperature evolution was reported for the A$'_1$ mode in monolayer MoS$_2$~\cite{MolasAcid}.
The temperature evolution of the infrared B$_\textrm{1u}$($\Gamma$) counterpart warrants attention as it displays a unique temperature dependence. 
A decrease in the intensity of the B$_\textrm{1u}$($\Gamma$) mode is observed up to about 150~K, which is followed by an almost constant intensity at higher temperatures. 
Surprisingly, the A$_\textrm{1g}$($\Gamma$) intensity is of about 5 times larger as compared to the B$_\textrm{1u}$($\Gamma$) one at $T$=5~K, then both intensities are almost equal at $T$=220~K, ending with about 3 times smaller intensity of A$_\textrm{1g}$($\Gamma$) mode as compared to the B$_\textrm{1u}$($\Gamma$) one.
One can conclude that at low temperature the main contribution for that Davydov pair originates from the Raman-active A$_\textrm{1g}$($\Gamma$) peak, while at room temperature the corresponding RS peak is dominated by the infrared-active B$_\textrm{1u}$($\Gamma$) mode.
The latter effect is different from the room temperature results on the flakes deposited on SiO$_{2}$/Si substrates, as the B$_\textrm{1u}$($\Gamma$) is hardly observed in the RS spectra for BLs at room temperature~\cite{Chakraborty2013, golasa2014resonant}.
This intensity switching between the A$_\textrm{1g}$($\Gamma$) and B$_\textrm{1u}$($\Gamma$) intensities is similar to the observation made from electric-field dependent RS in BL 2H-MoS$_{2}$~\cite{klein2021electrical}.
It was observed that the applied electric field provided enhanced electronic transitions from the dark Q valley to the bright K valley in the BZ due to breaking of crystal symmetry. 
This was because there is a strong delocalisation of the electron population in the conduction band at the high symmetry Q valley in both the layers and on application of an electric field, the formation of intralayer and interlayer excitons is facilitated by modifying the exciton population that leads to changes in the electron-phonon coupling. 
In our case, similarly, the strength of the electron-phonon coupling is modified by the variation of the temperature, which is discussed in the following.

In order to understand the effect of temperature on the phonon intensities, we investigate the intensity profiles of 6 other phonon modes, see Fig.~\ref{fig:fig3}(b) and (c). 
The temperature evolutions of the intensity profiles of TA+LA, 2LA, and E$_\textrm{1g}$ + TA peaks are very similar (see Fig.~\ref{fig:fig3}(b)). 
For the TA+LA mode, we observe a small increase in its intensity to about 150~K, which is followed by a quick exponential decay of almost 20 times in the temperature range from 150~K to 320~K. 
In the case of the 2LA and E$_\textrm{1g}$+TA modes, their intensities are almost constant to around 150~K and then they experience an analogous quick exponential decrease of about 15-20 times with increasing temperature to 320~K.
It should be noted that $T$=150 K corresponds to a relative energy of about 430~cm$^{-1}$, which is reasonably close to the energy of the TA+LA mode (429~cm$^{-1}$), while 2LA and E$_\textrm{1g}$+TA are at about 462~cm$^{-1}$ and 470~cm$^{-1}$, respectively.  
In our opinion, the similarity of the TA+LA energy and the relative energy of the 150~K is coincidental.
We also analyse the intensity evolution of the A$_\textrm{2u}$+TA, E$^1_\textrm{2g}$+LA and E$_\textrm{1g}$+2TA peaks as a function of temperature, shown in Fig.~\ref{fig:fig3}(c).
The intensities of those three lines follow the same pattern.
An increase in their intensities is observed up to about 180~K, which is followed by exponential decay at higher temperatures.
The temperature 180~K, corresponding to the relative energy of about 500~cm$^{-1}$, does not match the energies of the three peaks analyzed, which are in the range from 550 cm$^{-1}$ to 650 cm$^{-1}$.
This suggests that the effect of temperature on the resonant conditions of RS is more complex compared to the results of the RS excitation technique, where the laser energy is tuned in reference to the excitonic emission~\cite{Molas2017, Chow2017, Shree2018, Zinkiewicz2019}.

Summarising, the observed temperature effect on the intensity of the phonon modes is complicated and cannot be understood in terms of simple incoming or outgoing resonance conditions of the RS~\cite{Wu2018}.
In the presented experiment, the strength of the electron-phonon coupling is not only affected by the adjustment of the relative energy between the excitation and the emission (E$_{L-{X_{A}}}$). 
The variation in the temperature also alters the linewidth of the X$_{A}$ line, which is observed to broaden with increasing temperature, and hence the X$_{A}$ lifetime is changed~\cite{Palummo2015, Wang2016}.
The broadening of the A exciton line at higher temperatures results from the occupation by both electron and hole states characterised by larger $k$-vectors away from the K valleys.
As reported in \textcolor{red}{Ref.~}\citenum{Miranda2018} for thin MoTe$_2$ layers, the contributions to the Raman susceptibility from different BZ regions (individual $k$ points) are added with particular signs (plus or minus).
Consequently, the strength of the electron-phonon coupling can be significantly modified as a function of temperature.
This allows us to perform a quantitative analysis of the observed temperature dependent evolution of the phonon intensities but without direct attribution of resonant conditions of Raman scattering.

\section{Conclusion}
Temperature dependent resonant Raman spectroscopy for BL 2H-MoS$_{2}$ has been performed. 
It has been clearly observed that temperature plays a significant role in altering the band structure of the material, ultimately leading to a unique vibrational response from the active 2D layer, where different phonon modes at the $\Gamma$ and M points of the BZ are sensitive to temperature in an independent manner. 
The unambiguous switching of the intensity strength from A$_\textrm{1g}$($\Gamma$) at 5~K to B$_\textrm{1u}$($\Gamma$) at 300~K points to the fact that, in addition to the application of electric field, temperature is a crucial parameter to tune the resonance of the phonon modes in 2D materials.
We also note the effect of the flake environment on the Raman scattering.
We believe that our work will motivate further investigation from both an experimental and theoretical perspective on exciton-phonon coupling in S-TMD under resonant conditions of excitation. 

\section{Methods \label{methods}}
Bilayer MoS$_2$ encapsulated in hBN flakes was fabricated by a two-stage PDMS-based mechanical exfoliation of the bulk 2H-MoS$_{2}$ crystal. 
An unoxidised silicon wafer was used as a substrate. To ensure the best quality, the substrate was annealed at 200 $^o$C and kept on a hot plate until the first non-deterministic transfer of hBN flakes. 
Subsequent layers were transferred deterministically to reduce inhomogeneity between each layer. 
The complete structure was annealed at 160 $^o$C for 1.5 h to ensure the best layer-to-layer and layer-to-substrate adhesion and to eliminate air pockets at the interfaces between the constituent layers.

\textcolor{red}{A high resolution image of the sample was taken at 100x magnification using a Huvitz HRM-300 optical microscope. AFM experiments were conducted using Dimension Icon (Bruker Corporation, Billerica, MA, USA) with ScanAsyst connected to Nanoscope VI controller. The images were collected in ScanAsyst mode using SCANASYST-AIR probes (Bruker Corporation) across an area of 5~$\mu$m x 5~$\mu$m. with nominal spring constant of 0.4 N/m and a resonance frequency of 70 kHz. AFM topography images were recorded in air at the temperature of $23\pm1$~$^\circ$C.}

 \textcolor{red}{}

The PL and RS measurements were performed using $\lambda$ = 632.8 nm (1.96 eV) and $\lambda$ = 514.5 nm (2.41 eV) radiations from He-Ne and diode lasers, respectively. 
The laser beam, cleaned through Bragg filters in excitation, was focused through a 50x long-working-distance objective with a 0.55 numerical aperture producing a spot of about 1 $\mu$m diameter. 
The signal was collected via the same microscope objective, sent through a 1 m monochromator, and then detected by using a liquid nitrogen-cooled charge-coupled device (CCD) camera. 
The temperature-dependent PL and Raman measurements were performed by placing the sample on a cold finger in a continuous flow cryostat mounted on x–y motorised positioners. 
The temperature was varied from 5 K to 320 K in steps of 10 K with the signal collected from the sample after the stabilisation of the temperature. 
The excitation power focused on the sample was kept fixed at 300 $\mu$W during all measurements to obtain a strong signal and avoid local heating.

DFT calculations were conducted in Vienna Ab initio Simulation Package~\cite{VASP} with Projector Augmented Wave method~\cite{PAW}. Perdew–Burke–Ernzerhof parametrization~\cite{PBE} of general gradients approximation to the exchange-correlation functional was used. The plane waves basis cutoff energy was set to 500 eV and a 12$\times$12$\times$1 $\Gamma$-centered Monkhorst-Pack k-grid sampling was applied. The geometric structure was optimized with $10^{-5}$ eV/\AA and 0.01 kbar criteria for the interatomic forces and stress tensor components, respectively. Grimme's D3 correction was applied to describe the interlayer vdW interactions~\cite{D3}. The phonon band structure of BL MoS$_2$ was calculated within Parli\'nski-Li-Kawazoe method~\cite{Parlinski}, as implemented in Phonopy software~\cite{Phonopy}. The 3$\times$3$\times$1 supercells were found sufficient to converge the interatomic force constants within the harmonic approximation.

\section*{Acknowledgements \label{Acknowledgements}}
The work has been supported by the National Science Centre, Poland (grant no. 2017/27/B/ST3/00205 and 2018/31/B/ST3/02111) and the CNRS via IRP "2DM" project. K. W. and T. T. acknowledge support from the Elemental Strategy Initiative conducted by the MEXT, Japan, (grant no. JPMXP0112101001), JSPS KAKENHI (grant no. JP20H00354), and the CREST (JPMJCR15F3), JST. DFT calculations were performed with the support of the Interdisciplinary Centre for Mathematical and Computational Modelling (ICM) University of Warsaw and Center for Information Services and High Performance Computing (ZIH) at TU Dresden.

\section*{Conflicts of interest}
There are no conflicts to declare.

\section*{Availability of Data and Materials}
The datasets obtained during the experiments and analysed for the current study are available from the corresponding authors on reasonable request.

\bibliographystyle{apsrev4-1}
\bibliography{biblio}

\end{document}